\documentclass[12pt]{article}

\usepackage{amssymb}
\usepackage{amsfonts}
\usepackage{amsmath}
\usepackage{url}
\usepackage[margin=1.25in]{geometry}

\newtheorem{theorem}{Theorem}[section]

\title{Normalized Range Voting Broadly Resists Control\footnote{
    Supported in part by NSF grants CCF-0426761,
    IIS-0713061, CCF-0915792, and CCF-1101479. This work originally appeared as my
    master’s thesis at the Rochester Institute of Technology.  A
    version of this paper appears as University of Rochester technical
    report TR956.}}

\author{
Curtis Menton\\
Department of Computer Science \\
University of Rochester \\
Rochester, NY 14627, USA \\
menton@cs.rochester.edu
}

\begin{document}

\maketitle

\begin{abstract}
We study the behavior of Range Voting and Normalized Range Voting with
respect to electoral control. Electoral control encompasses attempts
from an election chair to alter the structure of an election in order
to change the outcome. We show that a voting system resists a case of
control by proving that performing that case of control is
computationally infeasible.  Range Voting is a natural extension of
approval voting, and Normalized Range Voting is a simple variant which
alters each vote to maximize the potential impact of each voter.  We
show that Normalized Range Voting has among the largest number of
control resistances among natural voting systems.
\end{abstract}


\section{Introduction}

Many of the key results in voting theory show that all voting systems
are flawed in some way.  Arrow's Impossibility Theorem states that in
any election with more than two candidates each voting system will
violate at least one of several reasonable and natural
criteria~\cite{arr:j:difficulty}.  The Gibbard--Satterthwaite Theorem
and Duggan--Schwartz Theorem show that all reasonable voting systems
are susceptible to strategic voting, where a voter may be able to vote
counter to his or her true preferences and achieve a better
outcome~\cite{gib:j:polsci:manipulation,sat:j:polsci:manipulation,dug-sch:j:polsci:gibbard}.
With any voting system, it might be possible for a dishonest election
organizer able to subtly alter the election to achieve his or her
desired end.  Thus much of the following study of voting systems has
been directed toward finding the best compromises and most reasonable,
if imperfect, solutions.

Electoral control represents cases where the authority conducting the
election attempts to alter the outcome by changing the structure of
the election.  The study of control of elections was initiated by
Bartholdi, Tovey, and Trick~\cite{bar-tov-tri:j:control}, who also
introduced a novel defense against it.  Even if control is possible,
it may be computationally very difficult to find an ideal plan.  The
standard tools of complexity theory can be brought to bear on the
problem.  In many cases, a control problem can be shown to be NP-hard
and thus very unlikely to be solvable in polynomial time.  We may be
able to accept theoretical vulnerability to control if computational
difficulty would make it essentially impossible for any
computationally limited attacker.

Since the initial work of Bartholdi, Tovey, and Trick, a number of
voting systems have been studied with an eye toward computational
resistance.  Several systems have been found with a high number of
resistances~\cite{erd-now-rot:j:spav,hem-hem-rot:j:hybrid,fal-hem-hem-rot:j:llull},
although some of them are not sufficiently natural for practical use,
remain vulnerable to some of the cases of control, or have other
technical flaws.  The system fallback voting was found by Erd{\'e}lyi
et al.~\cite{erd-pir-rot:c:buck-open} to be resistant to all but
two cases of control and it possesses the best known set of resistances for a
natural system. Voting systems have been developed that resist all
cases of control, but at the cost of being highly
unnatural~\cite{hem-hem-rot:j:hybrid}.  Thus it is still highly
desirable to search for natural and robust voting systems with high
degrees of control resistance.

This paper is particularly motivated by the work of Erd\'{e}lyi et
al.\ who studied the system sincere-strategy preference-based approval
voting (SP-AV)~\cite{erd-now-rot:j:spav}. This system is a hybrid of
approval voting and plurality, and it handily combines the control
resistances of both systems.  It does so by adding a vote coercion
step that adjusts all votes to approve of and disapprove of at least
one candidate.  This results in more complex behavior upon changes to
the candidate set and gives the system the candidate-control
resistances possessed by plurality.  However this may have the effect
of forcing a voter to distinguish between candidates he or she ranked
evenly and assigning him or her an arbitrary new approval threshold
that may not represent his or her preferences.

Range voting is a voting system with an alternate voter preference
representation that allows a voter to score his or her level of
approval of each candidate~\cite{smi:w:range1}.  Range voting has a
number of real world advocates due to it's good behavior regarding
conventional voting system criteria.  We will also introduce a variant
of range voting and show that it has among the highest degrees of
resistance to control among natural voting systems, matching the set
of control resistances possessed by Fallback voting.  This system,
normalized range voting, uses a similar vote-alteration procedure to
SP-AV, but in a way that preserves the relative preferences of a voter
among the candidates in the election.

\section{Range Voting}

\label{rv-ref}
Range voting (RV) is a voting system with an alternate voter
representation that allows voters to express his or her degree of
approval in each candidate.  Let $\|S\|$ denote the cardinality of a
finite set $S$.  We will describe a $k$-range election as
$E = (C,V)$ where $C$ is the set of candidates with $\|C\| = m$, and
$V$ is the set of voters with $\|V\| = n$ and for a voter $v \in V$,
$v \in \{0,1,\ldots,k\}^m$.  Each voter expresses his or her preferences by
giving a score for each candidate. The parameter $k$ sets the highest
score a voter is allowed to give a candidate.  The winners of the
election are the candidates with the highest sum score across all
voters.

\paragraph{Example}
The following is an example of a 2-range election of the candidates
$\{a,b,c\}$.  $a$ will be the winner with a total of 14 points.  

\begin{center}
\begin{tabular}{c|ccc}
  \# Voters & $a$ & $b$ & $c$ \\
\hline
  5 & 2 & 0 & 1 \\
  6 & 0 & 2 & 0 \\
  4 & 1 & 2 & 0
\end{tabular}
\end{center}

Though range voting is also sometimes described allowing scores over a real
interval such as $[0,1]$~\cite{smi:w:range1}, this paper will deal
with the more limited integral version for its practicality of
implementation and to avoid issues with the size of representation.
Our primary concern is to study the difficulty of decision problems
relating to the system and allowing scores of unbounded size would
greatly complicate that analysis.  Note that any bounded size and
precision real number representation would be equivalent to an
integral representation, so this version will be just as expressive as
a rational representation or any other which would be suitable for
computational analysis.

Arrow's Theorem was formulated with the traditional voter preference
models of a strict ordering.  Since range voting uses a different
model, it is not bound by that result and, though subject to
interpretation, achieves all of the criteria, which is normally
impossible for a voting system~\cite{smi:w:range1,hil:t:utilitarian}.

To demonstrate, we will show an example where RV satisfies
independence of irrelevant alternatives (IIA) while plurality would
violate it.  IIA is satisfied in an election system if the relative
ranking between two candidates is independent of the presence or
ranking of other candidates.  Let us formulate a 1-range
election, and assume that each voter only gives any points to his or
her top candidate.

\begin{center}
\begin{tabular}{c|ccc|c}
\# Voters & $a$ & $b$ & $c$ & Ranked ballot\\
\hline
5 & 1 & 0 & 0 & $a > b > c$\\
4 & 0 & 1 & 0 & $b > a > c$\\
2 & 0 & 0 & 1 & $c > b > a$
\end{tabular}
\end{center}

Candidate $a$ wins this initial election in either election system.
Now, if we remove the last place candidate $c$:

\begin{center}
\begin{tabular}{c|cc|c}
\# Voters & $a$ & $b$ & Ranked Ballot\\
\hline
5 & 1 & 0 & $a > b$\\
4 & 0 & 1 & $b > a$\\
2 & 0 & 0 & $b > a$
\end{tabular}
\end{center}

In the range election, the $c$ voters are left not awarding any points
to anybody, which is a perfectly legal vote, and perfectly rational,
if one does feel no distinction between the candidates.  Consequently
the original result stands and $a$ remains the winner.  In the
plurality election, the votes of third group of voters are transferred
to their second choice and $b$ becomes the new winner of the
election.

\section{Normalized Range Voting}
\label{nrv-ref}
A rational voter seeking to maximize his or her impact in an election would
always give his or her most preferred candidate the highest score possible
($k$) and his or her least preferred candidate the lowest score possible
($0$).  We introduce the system Normalized Range Voting (NRV), which
captures this and also gives the system more interesting behavior
under several types of centralized control.

In this system each voter specifies his or her preferences as in standard
range voting.  However, as part of the score aggregation, the system
normalizes each vote to a rational in the range $[0,k]$.  Formally, for a
voter $v$ and $v$'s maximum and minimum scores $m$ and $n$, his or her
score $s$ for a candidate is changed to $\frac{k(s - n)}{m-n}$.  If
$m=n$, a voter shows no preference among the candidates and this vote
will not be counted.  The system does not make an effort to coerce
such an unconcerned vote into one that distinguishes between the
candidates.

The relationship between RV and NRV is closely analogous to the
relationship between approval voting and SP-AV\@.  The normalization
step ends up removing several cases of control immunity, but it
introduces more complex behavior on alterations of the candidate set
that gain back a greater number of control resistances.

Unlike RV, NRV unambiguously fails the criteria independence of
irrelevant alternatives.  Consider a $2$-NRV election with $C =
\{a,b,c\}$ and $V$ below.

\begin{center}
\begin{tabular}{c|ccc}
\# Voters & $a$ & $b$ & $c$ \\
\hline
7 & 2 & 0 & 0 \\
4 & 0 & 2 & 0 \\
4 & 0 & 1 & 2 \\
\end{tabular}
\end{center}

The candidate~$a$ will win this election with a score of $14$, with $12$ and $8$ for
candidates $b$ and~$c$.  However, consider the election with the
same voters but with the candidate $c$ removed.

\begin{center}
\begin{tabular}{c|ccc}
\# Voters & $a$ & $b$ \\
\hline
7 & 2 & 0 \\
4 & 0 & 2 \\
4 & 0 & 1 \\
\end{tabular}
\end{center}

At first, $a$ appears to still be winning the election.  However the
normalization step will scale up the votes from the third group of
voters to give $b$ 16 points in total, making $b$ the winner of the
election.  

While this seems to be a negative against this system, this complex,
shifting behavior on the changing of the candidates is exactly what
allows us to achieve a large number of control resistances for NRV
over RV\@.

\section{Control}

Control represents the efforts of a centralized authority, the chair
of an election to alter the structure of the election in order to
affect its outcome.  This involves changing either the candidate or
voter sets or partitioning either into subelections.  In real
world political elections, this corresponds to voter fraud and voter
suppression, back-room dealings with potential candidates, and
gerrymandering and similar manipulations.  In the context of
multiagent systems, it is related to any efforts by a system designer
or administrator to alter the results by changing the parameters of
the system.

More formally, for the purposes of the complexity theoretic analysis
of the control problems, we will analyze the cases of control in the
form of decision problems.  That is, we will define a problem where
the goal is to find whether in a particular election a certain case of
control can succeed in its goals.

The goal is to classify a voting system as vulnerable, resistant, or
immune to each of the various cases of control, with these terms
initially defined by Bartholdi et al.\ \cite{bar-tov-tri:j:control} and widely
adopted since.  It is helpful now to
define these notions precisely.

\paragraph{Vulnerability}
A voting system is \emph{vulnerable} to a case of control if that
action has potential to affect the result of an election, and the
associated decision problem can be solved in polynomial time; that is,
it is in P.  This has a very good practical correspondence with real
world efficiency of the problem, and thus the case of control is
computationally easy.

\paragraph{Resistance}
A voting system is \emph{resistant} to a case of control if that
action has potential to affect the result of an election, and the
associated decision problem is NP-hard.  The idea of NP-hardness has a
long and storied history, but for the current purposes, it suffices to
say that such problems are very unlikely to have efficient solutions,
barring a major shift in our understanding of computer science.

\paragraph{Immunity}
A voting system is \emph{immune} to a case of control if that action
cannot affect the result of the election.  This is obviously a
desirable notion but it is generally harder to come by, so often we
must be satisfied with computational resistance instead.

The control cases of Bartholdi et al.\ were all \emph{constructive},
that is, the control is directed towards making a distinguished
candidate the winner.  In some cases, a malicious chair  could conceivably
want above all to prevent a particular candidate from winning the
election, regardless of who else wins.  This idea was introduced by
Conitzer et al.\ as \emph{destructive} manipulation and later by
Hemaspaandra et al.\ in the context of control ~\cite{hem-hem-rot:j:destructive-control}.
Though this may seem to be a less desirable goal, it may be feasible
in some cases where constructive control is not and thus is it also
worth studying.

Among the cases of control are control by adding or deleting either
voters or candidates.  In the case of adding voters or candidates, the
new participants must be chosen from a set rather than arbitrarily
created.  The decision problems are defined here as having a limit on
the number of voters or candidates that can be added or deleted,
though we also include the case of unlimited adding of candidates, as
this was the version used in the original paper on
control~\cite{bar-tov-tri:j:control} and including this extra case has
become standard in the literature.  In the candidate cases, the
distinguished candidate must be in the original candidate set.  In the
cases of destructive control by deleting candidates, the distinguished
candidate cannot be among those deleted as that would trivially solve
the problem.

The various cases of control by partition are not quite
straightforward and deserve a little explanation.  In any control by
partition problem, initial subelections are performed with segments of
the voter and candidate sets and a final election is performed with
the candidates that survive these subelections.

In control by partition of voters, the voter set is partitioned into
two subsets and an subelections are run with each (with the original
candidate set).  The candidates that survive each subelection face off
to find the final winner of the election.  

Control by partition of candidates has two major variants.  In one
variant, \emph{control by partition}, one set of candidates is
separated off from the rest for an initial subelection.  Whatever
candidates survive this election then rejoin the rest of the
candidates for the final election with the entire voter set.  In the
other variant, \emph{control by run-off partition}, the candidate set
is partitioned into two sets and each set conducts an initial
subelection.  The candidates that survive each of these elections then
are brought together for the final election with the entire voter set.

There is an additional variation in the tiebreaking rule that is
chosen in the subelections.  In the case of a tie, either all of the
top scoring candidates are promoted to the final election, or none of
them are.  These two cases are called \emph{ties-promote} and
\emph{ties-eliminate}.  Notably, in the second case, an election can
fail to elect any candidate.  Though these may seem like subtle
differences, many voting systems will resist one of the cases while
being vulnerable to another.  

We use in this paper the unique-winner model, following the original paper
on control~\cite{bar-tov-tri:j:control}.  Here, the goal in the
control problems in the constructive cases is to find whether we can
make our preferred candidate a unique winner, so they must both be a
winner of the election and the only winner.  In the destructive cases,
in a successful instance, there must either be multiple winners, or a
single winner that is not the hated candidate, or no winners.

\subsection*{Control by Adding Candidates}
\begin{description}
\item[Given] An election $E=(C,V)$, a distinguished candidate $w \in C$, a
  spoiler candidate set $D$, and $k \in \mathbb{N}$.
\item[Question (Constructive] Is it possible to make $w$ a unique winner of an election
  $(C \cup D', V)$ with some $D' \subseteq D$ where $|D'| \leq k$?

\item[Question (Destructive)] Is it possible to make $w$ not a unique winner of an election
  $(C \cup D', V)$ with some $D' \subseteq D$ where $|D'| \leq k$?
\end{description}

\subsection*{Control by Adding an Unlimited Number of Candidates}
\begin{description}
\item[Given] An election $E=(C,V)$, a distinguished candidate $w \in
  C$, and a spoiler candidate set $D$.
\item[Question (Constructive] Is it possible to make $w$ a unique winner of an election
  $(C \cup D', V)$ with some $D' \subseteq D$?

\item[Question (Destructive)] Is it possible to make $w$ not a unique  winner of an election
  $(C \cup D', V)$ with some $D' \subseteq D$?
\end{description}

\subsection*{Control by Deleting Candidates}
\begin{description}
\item[Given] An election $E=(C,V)$, a distinguished candidate $w \in
  C$, and $k \in \mathbb{N}$.
\item[Question (Constructive)] Is it possible to make $w$ a unique winner of an election
  $(C-C', V)$ with some $C' \subseteq C$ where $|C'| \leq k$?

\item[Question (Destructive)] Is it possible to make $w$ not a unique  winner of an election
  $(C-C', V)$ with some $C' \subseteq (C - \{w\})$ where $|C'| \leq k$?
\end{description}

\subsection*{Control by Adding Voters}
\begin{description}
\item[Given] An election $E=(C,V)$, a distinguished candidate $w \in
  C$, an additional voter set $U$, and $k \in \mathbb{N}$.
\item[Question (Constructive)] Is it possible to make $w$ a unique  winner of an
  election $(C, V \cup U')$ for some $U' \subseteq U$ where $|U'| \leq
  k$?

\item[Question (Destructive)] Is it possible to make $w$ not a unique winner of an
  election $(C, V \cup U')$ for some $U' \subseteq U$ where $|U'| \leq
  k$?
\end{description}

\subsection*{Control by Deleting Voters}
\begin{description}
\item[Given] An election $E=(C,V)$, a distinguished candidate $w \in
  C$, and $k \in \mathbb{N}$.
\item[Question (Constructive)] Is it possible to make $w$ a unique  winner of an
  election $(C, V - V')$ for some $V' \subseteq V$ where $|V'| \leq
  k$?
\item[Question (Destructive)] Is it possible to make $w$ not a unique winner of an
  election $(C, V - V')$ for some $V' \subseteq V$ where $|V'| \leq
  k$?
\end{description}

\subsection*{Control by Partition of Candidates}
\begin{description}
\item[Given] An election $E=(C,V)$ and a distinguished candidate $w
  \in C$.
\item[Question (Constructive)] Is there a partition $C_1, C_2$ of $C$
  such that $w$ is a unique final winner of the election $(D \cup C_2, V)$,
  where $D$ is the set of candidates surviving the initial subelection
  $(C_1, V)$?

\item[Question (Destructive)] Is there a partition $C_1, C_2$ of $C$
  such that $w$ is not a unique final winner of the election $(D \cup C_2,
  V)$, where $D$ is the set of candidates surviving the subelection
  $(C_1, V)$?
\end{description}

\subsection*{Control by Runoff Partition of Candidates}
\begin{description}
\item[Given] An election $E=(C,V)$ and a distinguished candidate $w
  \in C$.
\item[Question (Constructive)] Is there a partition $C_1, C_2$ of $C$ such that $w$
  is a unique final winner of the election $(D_1 \cup D_2, V)$, where $D_1$
  and $D_2$ are the sets of surviving candidates from the subelections
  $(C_1, V)$ and $(C_2, V)$?

\item[Question (Destructive)] Is there a partition $C_1, C_2$ of $C$ such that $w$
  is a unique final winner of the election $(D_1 \cup D_2, V)$, where $D_1$
  and $D_2$ are the sets of surviving candidates from the subelections
  $(C_1, V)$ and $(C_2, V)$?
\end{description}

\subsection*{Control by Partition of Voters}
\begin{description}
\item[Given] An election $E=(C,V)$ and a distinguished candidate $w
  \in C$.
\item[Question (Constructive)] Is there a partition $V_1, V_2$ of $V$ such that $w$
  is a unique  final winner of the election $(D_1 \cup D_2, V)$ where $D_1$
  and $D_2$ are the sets of surviving
  candidates from the subelections $(C, V_1)$ and $(C, V_2)$?

\item[Question (Destructive)] Is there a partition $V_1, V_2$ of $V$ such that $w$
  is not a unique  final winner of the election $(D_1 \cup D_2, V)$ where $D_1$
  and $D_2$ are the sets of surviving
  candidates from the subelections $(C, V_1)$ and $(C, V_2)$?
\end{description}

\section{Results}

The control results for these two systems, as well as approval and
SP-AV for comparison, are summarized in Table 1. ``V'', ``I'', and
``R'' stand for vulnerable, immune, and resistant, which are used in
the standard way in the literature dating from Bartholdi, Tovey, and
Trick~\cite{bar-tov-tri:j:control}. ``C'' marks
constructive control while ``D'' marks destructive control.  


\begin{table*}[htp]
  \footnotesize
  \centering
  \label {RV-NRV-results}
  \begin{tabular}{|l|l||l|l|l|l|l|l|l|l|l|l|}
    \hline
    Control by & Tie & \multicolumn{2}{|c|}{Approval} &
    \multicolumn{2}{|c|}{SPAV} & \multicolumn{2}{|c|}{Fallback} &  \multicolumn{2}{|c|}{RV} &
    \multicolumn{2}{|c|}{NRV} \\
    & Model & C& D & C& D & C& D & C& D & C& D\\
    \hline \hline
    Adding Candidates & & I & V & R & R & R & R & I & V & R & R\\
    \hline
    Adding an Unlim. Number of Candidates & & I & V & R & R & R & R & I & V & R & R\\
    \hline
    Deleting candidates & & V & I & R & R & R & R & V & I & R & R\\
    \hline
    Partition of Candidates & TE & V & I & R & R & R & R & V & I & R & R\\
    \hline
    & TP & I & I & R & R & R & R & I & I & R & R \\
    \hline
    Run-off Partition of Candidates & TE & V & I & R & R & R & R & V & I & R & R\\
    \hline
    & TP & I & I & R & R & R & R & I & I & R & R\\
    \hline
    Adding Voters & & R & V & R & V & R & V & R & V & R & V\\
    \hline
    Deleting Voters & & R & V & R & V & R & V & R & V & R & V\\
    \hline
    Partition of Voters & TE & R & V & R & V & R & R & R & V & R & R\\
    \hline
    & TP & R & V & R & R & R & R & R & V & R & R \\
    \hline
  \end{tabular}
  \caption{Control Results for Approval, Sincere-Strategy
    Preference-Based Approval Voting~\cite{erd-now-rot:j:spav},
    Fallback Voting~\cite{erd-pir-rot:c:buck-open}, Range Voting, Normalized Range
    Voting}
\end{table*}

\subsection{Generalization of Resistance Results}

The proofs here will refer necessarily to specific RV and NRV
elections with a particular scoring range $k$, and additionally they
will use different values as necessary.  However we want to be able to
show resistance for other scoring ranges, and show that all the
resistances we show will hold for some particular scoring range.

\begin{theorem}
If RV or NRV exhibits resistance to a case of control for a particular
scoring range $k$, it will exhibit that resistance for any range $ak$
with $a \in {\mathbb N}^+$.
\end{theorem}

\paragraph{Proof} 
We can reduce an instance of any RV or NRV control problem for an
election with a scoring range $k$ to an instance of the same problem
with an election with a scoring range of $ak$ for any $a \in {\mathbb
  N}^+$.  We can do this simply by scaling all of the scores in all of
the
votes in the original election up by a factor of $a$.  This new
election with the new scoring bound and new votes will behave the same
as the original election before any control attempt, and it will also
behave the same under any control attempt or manipulative action, as
all of the votes and the sum scores will be scaled up by the same
factor $a$.  In the case of NRV, any normalization that occurs in the
original election will occur in the newly scaled election to the same
degree, but just with the pre and post normalization scores both being
scaled up by the factor of $a$. Thus the winner in the scaled election
will be the same before and after any control action and control
problems easily reduce to same problem in the scaled voting system. $\Box$

\subsection{Results Derived From Approval}

Due to RV's great similarity with approval voting, many results
relating to approval trivially apply to RV and NRV\@.

\begin{theorem}
If approval voting is resistant to a case of control, RV and NRV will
also be resistant for any scoring range.  
\end{theorem}

This is easy to show.  We can reduce from an instance of any approval
control problem by simply considering the election a 1-range election
or a 1-normalized-range election. A 1-range election is exactly
equivalent so this will trivially work.  For the NRV election, though
this does technically include the normalization step which can modify
the election, when the score range is 1, no normalization is actually
performed, so again this election is equivalent to the original
approval election.  These results will also generalize to $k$-RV and
$k$-NRV any $k \geq 1$ as previously described.

\begin{theorem}
1-RV and 1-NRV are resistant to the following cases of control:
constructive control by adding voters, constructive control by
deleting voters, and constructive control by the partition of voters in
the ties-promote and ties-eliminate models.
\end{theorem}

All of these resistances are derived from reductions from approval as
described above, and the fact that approval is resistant to these
cases of control~\cite{hem-hem-rot:j:destructive-control}.

\subsection{Adding/Deleting Candidates}

\begin{theorem}
2-NRV is resistant to constructive control and destructive control by
adding candidates, constructive and destructive control by adding an
unlimited number of candidates, and destructive control by deleting
candidates.
\end{theorem}

\paragraph{Proof.}
This proof is inspired by a similar proof relating in SP-AV by
Erd{\'e}lyi, Nowak, and Rothe~\cite{erd-now-rot:j:spav}.

We will reduce from an instance of the hitting set problem, defined as
follows~\cite{gar-joh:b:int}.

\begin{description}
\item[Given:] A collection ${\cal S}$ of subsets of a set $B$, $k \in \mathbb{N}^+$
\item[Question:] Does $B$ contain a hitting set $B'$ of size $k$ or
  less that contains at least one element from every $S \in {\cal S}$?
\end{description}

Given a hitting set instance $(B, (\cal S), k)$ with $|B| = n$ and
$|{\cal s}| = m$ we will construct a 2-range election.  The candidate
set $C$ will consist of $B \cup \{c,w\}$.  The idea is that $c$ will
win the election unless only a hitting set of size $k$ of candidates
from $B$ are included.  The voter set $V$ will be as follows:

\begin{itemize}
\item $2m(k+1) + 4n$ voters have a score of 2 for $c$, and a score of
  0 for all other candidates.
\item $3m(k+1) + 2k + 1$ voters have a score of 2 for $w$, and a score
  of 0 for all other candidates.
\item For each $b \in B$, 4 voters have a score of 2 for $b$, a score
  of 1 for $w$, and a score of 0 for all other candidates.
\item For each $S_i \in {\cal S}$, $2(k+1)$ voters have a score of 2 for $b$, for
  each $b \in S_i$, a score of 1 for $c$, and a score of 0 for all other candidates.
\end{itemize}

This will lead to scores in $(\{c,w\}, V)$ as follows:

\begin{center}
\begin{tabular}{c|c}
  Candidate  & Score \\
  \hline
  $c$ & $8m(k+1) + 8n$\\
  $w$ & $6m(k+1) + 8n + 4k + 2$
\end{tabular}.
\end{center}

The candidate $c$ will win with a margin of $2m(k+1) - 4k - 2$.

Additionally the scores in $(\{c,w\} \cup B, V)$ will be as follows:

\begin{center}
\begin{tabular}{c|c}
  Candidate  & Score \\
  \hline
  $c$ & $6m(k+1) + 8n$\\
  $w$ & $6m(k+1) + 4n + 4k + 2$\\
  $b \in B$ & $\leq 8 + 4m(k+1) $ 
\end{tabular}.
\end{center}

Here, $c$ will win with a margin of $4n - 4k - 2$, which will be
positive as long as $k < n$.



We will show that $w$ will be the winner of $(\{c,w\} \cup B', V)$
with $B' \subseteq B$ if $B'$ corresponds to a hitting set of size
$\leq k$.  The candidate $w$ loses 4 points for each $b \in B'$
included, of which there are no more than $k$.  $c$ loses $2(k+1)$
points for each $S_i$ hit.  There will be $m$ such sets if $B'$ is a
hitting set, so $c$ loses $2m(k+1)$ points total.

\begin{center}
\begin{tabular}{c|c}
  Candidate  & Score \\
  \hline
  $c$ & $6m(k+1) + 8n$\\
  $w$ &  $ \geq 6m(k+1) + 8n + 2$\\
  $b \in B'$ & $\leq 8 + 4m(k+1) $ \\
\end{tabular}
\end{center}

$w$ will end up with an advantage of at least 2 points and thus
$w$ will be the winner of the election.

We will show that $w$ will not be the winner of any election $(\{c,w\}
\cup B', V)$ where $B'$ does not correspond to a hitting set of size
$\leq k$.
If $B'$ is a hitting
set but $|B'| > k$, $c$ will have $6m(k+1) + 8n$ points and $w$ will
have $\leq 6m(k+1) + 8n$.  If $B'$ is not a hitting set $c$ will have
$\geq 6m(k+1) + 8n + 2(k+1)$ points and $w$ will have $\leq 6m(k+1) +
8n + 2k + 2$ points.  In either case $w$ will not be the unique
winner.

This construction can be used to create cases of constructive and
destructive control by adding candidates and destructive control by
deleting candidates.  $((C,V), (m-k), c)$ is such an instance of
destructive control by deleting candidates.  $((\{c,w\}, V), B, k, w)$
is an instance of constructive control by adding candidates, and
$((\{c,w\}, V), B, k, c)$ is an instance of destructive control by
adding candidates. $((\{c,w\}, V), B, w)$ and $((\{c,w\}, V), B, c)$
are appropriate instances of constructive and destructive control by
adding an unlimited number of candidates. $\Box$

\begin{theorem}
2-NRV is resistant to constructive control by deleting candidates.
\end{theorem}

As in Erd\'{e}lyi, Nowak, and Rothe~\cite{erd-now-rot:j:spav}, the previous reduction is not
sufficient to show resistance to constructive control by deleting
candidates, as $c$ and $w$ are the only candidates with a shot at
winning.  Deleting $c$ will instantly make $w$ the winner.  The
remaining case can be handled by the following reduction.

\paragraph{Proof}
We will reduce from an instance of hitting set $(B, {\cal S},
k)$.

The candidate set $C$ will consist of $B \cup \{w\}$, the set $B$ from
the instance of hitting set together with an additional candidate.

The voter set $V$ will be constructed as follows:
\begin{itemize}
\item 
  $n+k$ voters have a score of 2 for $b$ for every $b \in B$, and a
  score of 0 for $w$.
\item
  $3 + 2mk$ voters have a score of 2 for $w$ and a score of 0 for all other
  candidates.
\item 
  For each $S \in {\cal S}$, $4k+1$ voters have a score of 2 for $s$ for
  every $s \in S$, a score of 1 for each candidate in $B - S$, and
  a score of 0 for $w$.
\item 
  For each $S \in {\cal S}$, $4k+1$ voters have a score of 2 for $b$ for
  every $b \in B - S$, a score of 2 for $w$, and a score of 1 for $s$
  for every $s \in S$
\item
  For each $b \in B$, $2n-k$ voters have a score of 2 for $b$, a
  score of 1 for $w$, and a score of 0 for every other candidate.
\end{itemize}


We can show that if $B'$ is a hitting set of size $k$, $w$ will be the
winner of the election $(B'\cup\{w\},V)$.
Assume $B'$ is a hitting set and $\|B'\| = k$. Each $b \in B'$ will
receive $12mk + 4n - 2k + 4$ points.  $w$ will receive $8mk + 6
+ 4mk + 4(n - k) + 2k = 12mk + 4n - 2k + 6$ points.  Thus
$w$ will be the winner in the election.

We can show that if $B'$ is not a hitting set or if $\|B'\| > k$, $w$
will not be the winner of the election $(B'\cup\{w\},V)$.
First assume $B'$ is a hitting set but $\|B'\| = l > k$.  Since $B'$ is
a hitting set, every $b \in B'$ will receive exactly $12mk + 4n - 2k +
4$ points.  $w$ will receive $4mk + 6 + 8mk + 4(n - l) + 2l = 12mk +
4n - 2l + 4$ points.  $score(b) - score(w) = -2k + 2l - 2$ which is
non-negative since $l > k$.  Thus $w$ will lose the election.

Next consider the case where $\|B'\| = l \leq k$ but $B'$ is not a
hitting set.  Thus every $b \in B'$ will have a score $ \geq 12nk
+ 4m + 2k + 4$ as each will gain an extra $4k$ points from one
set of group 3 voters.  $w$ will have the score $12nk + 4m - 2l + 6$.
Thus $score(b) - score(w) = 2k + 2l - 2$ which is non-negative
and $w$ will again lose the election.

An instance of hitting set $(B, {\cal S}, k)$ can thus be reduced to
finding whether $w$ can be made the winner of $(C,V)$ as above by
deleting $m-k$ candidates.~$\Box$

\subsection{Destructive Control by Partition of Voters}

\begin{theorem}
2-NRV is resistant to destructive control by partition of voters in the
ties-promote model.
\end{theorem}

\paragraph{Proof}

We will reduce from restricted hitting set.  Restricted hitting set is
an NP-complete hitting set variant introduced by Hemaspaandra,
Hemaspaandra, and Rothe with additional restrictions on the sizes of
the sets in an instance~\cite{hem-hem-rot:j:destructive-control}.  The
version as used here has a slightly stronger bound that is necessary
due to the somewhat larger numbers required in this proof.

\begin{description}
\item[Given:] A collection ${\cal S}$ of subsets of a set $B$, $k \in
  \mathbb{N}^+$, with $|{\cal S}| = m$, $|B| = n$, and the additional
    restriction that $m(k+1) +3 \leq n - k$
\item[Question:] Does $B$ contain a hitting set $B'$ of size $k$ or
  less that contains at least one element from every $S \in {\cal S}$?
\end{description}

Given an instance of restricted hitting set $(B, {\cal S}, k)$ with
$|B| = n$ and $|{\cal S}| = m$, create a 2-normalized-range election
with $C = B \cup \{w,c\}$ and $V$ as follows.

\begin{itemize}
\item $2m(k+1) + 4n$ voters have a score of 2 for $c$, and a score of
  0 for every other candidate.
\item $3m(k+1) + 2k$ voters have a score of 2 for $w$, and a score of
  0 for every other candidate.
\item For each $b \in B$, 4 voters have a score of 2 for $b$, a score
  of 1 for $w$, and a score of 0 for every other candidate.
\item For each $S_i \in {\cal S}$, for each $b \in S_i$, $2(k+1)$
  voters have a score of 2 for $b$, a score of 1 for $c$, and a score
  of 0 for every other candidate.
\item For each $b \in B$, 1 voter has a score of 2 for $b$ and a score of
  0 for every other candidate.
\end{itemize}

The candidate $c$ can be made to lose $(C,V)$ through partition of
voters if and only if there is a hitting set of size $\leq k$ over
${\cal S}$ in $B$.

We can show that if there is a hitting set of size $\leq k$, it is
possible to cause $c$ to lose the election through partition of
voters.  Given an appropriate hitting set $B'$, partition $V$ into
sets $V_1$ and $V_2$. Let $V_1$ contain a voter from the final group
corresponding to every $b \in B'$ and one voter from the second group
(allotting just 2 points to $w$) and let $V_2 = V - V_1$.  After the
initial subelections, we will be left with $w$, $c$, and the
candidates $B'$ corresponding to the hitting set, and $w$ will win
this election (see the reduction to adding/deleting candidates for the
details of that proof).

If there is no hitting set $B' \in B$ of size $\leq k$, $c$ cannot be
made to lose the election through partition of voters.  For any
actions attempting to control the election by forcing the final
candidate set, see the previous reduction to adding/deleting
candidates.  As for other efforts concentrated at more typically
partitioning the voters, among the initial candidates, $c$ has as high
of a score as any two other candidates, so $c$ must at least tie in
at least one of the subelections. Thus he or she will always make it to
the final election.  The scores of the candidates in the initial
election follow.

\begin{center}
\begin{tabular}{c|c}
Candidate & Score \\
\hline
$c$ &  $6m(k+1) + 8n$ \\
$w$ &  $6m(k+1) + 4k + 4n$ \\
$b$ & $\leq 4m(k+1) + 10$ \\
\end{tabular}
\end{center}

$c$'s score minus the next two highest scores will thus be at
least $4(n - k) - 4m(k+1) - 10$.  However, due to our use of
restricted hitting set, we have that $m(k+1) +3 \leq n - k$, and so
this is at least $2$. Thus the only way to defeat $c$ is to face
them against a hitting set of candidates as described. $\Box$


\begin{theorem}
4-NRV is resistant to destructive control by partition of voters in
the ties-eliminate model.
\end{theorem}

We will reduce from the Exact Cover by Three-sets problem
(X3C), defined as follows.

\begin{description}
\item[GIVEN] A set $B = \{b_1, \ldots, b_{3k}\}$ and a family ${\cal
  S} = \{S_1, \dots, S_n\}$ of sets of size three of elements from $B$.
\item[QUESTION] Is it possible to select $k$ sets from ${\cal S}$
  such that their union is exactly $B$?
\end{description}

\noindent
\emph{Proof.}

Given a X3C instance $B, {\cal S}$ we will construct a 2-range
election $(C,V)$ as follows.  The candidate set will be $B \cup
\{c,w\}$, where $c$ will be the distinguished candidate.  The voters
set $V$ will consist of the following:

\begin{itemize}
\item 
  For every $S_i \in {\cal S}$, one voter with a score of $4$ for
  every candidate in $B - S_i$, a score of $2$ for $c$, and a score of
  $0$ for every other candidate;
\item 
  $2n$ voters with a score of $4$ for every candidate in $B$, a
  score of $2$ for $c$, and a score of $0$ for $w$;
\item 
  $k-1$ voters with a score of $4$ for $w$, a score of $2$ for
  $c$, and a score of 0 for all other candidates;
\item 
  For every $b \in B$, $1$ voter with a score of $4$ for $b$, a
  score of $1$ for every candidate in $B-b$, a score of $1$ for $c$,
  and a score of 0 for $w$
\item 
  $2k+3n+1$ voters with a score of $4$ for $w$ and a score of $0$
  for every other candidate.
\end{itemize}

We will assume that $1 \leq  k \leq n$ and that each element in $B$ is in
at least one set from ${\cal S}$.  

If there is an exact cover over $B$, then $w$ can be made to lose the
election through partition of voters in the ties-eliminate model.
Consider the partition of the voter set into $V_1, V_2$, where $V_1$
consists of the voters from the first group corresponding to the
elements of the set cover together with the third group of voters, and
where $V_2 = V - V_1$.

The candidate $w$ will win the subelection $(C,V_2)$ and the
distinguished candidate $c$ will easily win the subelection $(C,
V_1)$.  $c$  will receive $2k$ points from the cover voters and $2k-2$
points from the other voters for a total of $4k-2$ points.  For the
$B$ candidates, each will receive $4$ points from each of the
first-group voters except the one that corresponds to the set that
covers them, coming to $4k-4$ points in total.  $w$ will gain $4$
points from each of the $k-1$ voters that favors them, and so he or she
will gain $4k-4$ votes in total.  Therefore $c$ will win this
subelection and will go on to face $w$ in the final election.

In the final election, with almost all of the candidates eliminated,
vote normalization will occur benefiting $c$ to the point that he or she
gains the advantage.  All of the votes in the first, second, and fourth groups
will now give four points to $c$, giving them $12n + 14k - 2$ points in
total.  This is at least as many as the $12n + 12k$ points $w$ was given,
and so $w$ will no longer be the unique winner.

If there is no exact cover, $w$ cannot be made to lose the election
through partition of voters in the ties-eliminate model.  This is due
to several facts: $w$ will win at least one of the two subelections,
$c$ cannot win either subelection, and $w$ will win head to head
against all candidate that it may face in the final election.

$c$ cannot win an  initial subelection except as previously described.
No voter prefers them outright, so the only way to make them win is to
balance the points he or she gains from the first group of voters and from
the third group of voters and to give he or she an advantage over each $B$
candidate by covering that candidate with a voter that prefers $c$.
None of the other voters will help $c$ win a subelection, as no others
help $c$ gain points relative to the $B$ candidates.  If a set of
first group voters smaller than $k$ that is not a cover is chosen,
then at least one $B$ candidate will gain points for every one of
these voters.  We will then not be able to boost $c$ over the $B$
candidates without giving too many points to $w$.  If a cover larger
than $k$ of voters is chosen, then there will not be enough third
group voters to include to boost $c$ over the $B$ voters and $c$ will
not be able to win.

The candidate $w$ must win at least one of the initial subelections
and make it to the final election.  With the original candidate set
and unnormalized scores, $w$ has a considerable advantage in points
over all other candidates, and since it is largely the same voters that
support all of the other candidates, there is no way to partition the
voters to make $w$ lose both subelections.

Against all candidates but $c$, $w$ will win a head to head
contest in the final election.  In a head-to-head contest with $w$ the
other candidates gain about as many points as $c$ through
normalization, but they will still have fewer points total as each $B$
candidate loses out on 4 points for every subset $S_i$ they are a part
of.  The $B$ candidate will thus have no more than $12n + 12k - 4$
votes, while $w$ will have $12n + 12k$ votes and will be the winner.

Therefore $w$ will win the final election for all partitions in the
case that there is not an exact cover over the set $B$.  $\Box$

\subsection{Partition of Candidates}

\begin{theorem}
4-NRV is resistant to constructive control by partition and runoff
partition of candidates.
\end{theorem}

We will reduce from control by deletion of candidates in NRV\@.  We will
show the reduction to constructive control by run-off partition,
though the other partition case is quite similar.

\paragraph{Proof}
Given an $r$-NRV election\footnote{Note that since $k$-NRV is
  resistant to deletion of candidates for $k \geq 2$, this reduction
  shows resistance for $k \geq 4$.}
 $(C,V)$ with $|C| = m$ and $|V|=n$
the distinguished candidate $w \in C$, and a deletion limit $k \in
\mathbb{N}$, construct a $2r$-NRV election $(C', V')$ as follows.  $C'
= C \cup \{a,b\}$, where $a$ and $b$ are additional auxiliary
candidates.  $V'$ will consist of the original voter set $V$ in
addition to the following voters.

\begin{itemize}
\item For each $c \in C$, $2n$ voters have a score of $2r$ for $c$, a
  score of $r$ for $a$, and a score of 0 for every other candidate.
\item For each $c \in C^*$, $3nm$ voters have a score of $2r$ for $c$
  and a score of 0 for every other candidate.
\item $2nm$ voters have a score of $2r$ for $w$, a score of $r$ for
  $a$, and a score of 0 for every other candidate.
\item $nm$ voters have a score of $2r$ for $w$ and a score of 0 for
  every other candidate.
\item $(m-k-1)n$ voters have a score of $2r$ for all $c \in C$ and a
  score of 0 for every other candidate.
\item $2n + 1$ voters have a score of $2r$ for $a$ and a score of 0
  for every other candidate.
\item $3n + 3nm + (m-k-1)n + 2$ voters have a score of $2r$ for b and a score of 0 for every other candidate.
\end{itemize}

Let $s_0(c)$ be the score of candidate $c$ among the original voters
$V$.  Note for any candidate $s_0(c) \leq nr$.

The following are the scores for the candidates in $(C', V')$ and various relevant
subelections thereof.

\vspace{1em}

\noindent
$(C', V')$

\begin{center}
\begin{tabular}{c|c}
  Candidate & Score \\
  \hline
  $a$         & $4nmr + 4nr + 2r$ \\
  $b$         & $6nr + 6nmr + 2(m-k-1)nr + 4r$ \\
  $c \in C^*$ & $4nr + 6nmr  + 2(m-k-1)nr + 2 s_0(c)$\\
  $w$         & $4nr + 6nmr + 2(m-k-1)nr + 2 s_0(w)$\\
\end{tabular}
\end{center}

The winner in this case will be $b$.

\vspace{1em}

\noindent
$(\{a,w\},V')$

\begin{center}
\begin{tabular}{c|c}
  Candidate & Score \\
  \hline
  $a$ & $4nr + 6nmr + 2$\\
  $w$ & $4nr + 6nmr + 2(m-k-1)nr + 2 s_0(w)$\\
\end{tabular}
\end{center}

The winner of this election will be $w$.

\vspace{1em}

\noindent
$(\{w\} \cup D, V')$ where $D \subseteq C^*, |D| = l$

\begin{center}
\begin{tabular}{c|c}
  Candidate & Score \\
  \hline
  $c \in D$ & $4nr + 6nmr + 2(m-k-1)nr + 2 s_0(c)$ \\
  $w$ & $4nr + 6nmr + 2(m-k-1)nr + 2 s_0(w)$\\
\end{tabular}
\end{center}

The winner will again be whatever candidate is the winner over the
original voter set $V$.

\vspace{1em}

\noindent
$(\{a,b\} \cup D, V')$ where $D \subseteq C^*$, $|D| = l$

\begin{center}
\begin{tabular}{c|c}
  Candidate & Score \\
  \hline
  $a$         & $4nr + 6nmr + 2(m-l)nr + 2r$\\
  $b$         & $6nr + 6nmr + 2(m-k-1)nr + 4r$ \\
  $c \in D$   & $4nr + 6nmr + 2(m-k-1)nr + 2 s_0(c)$\\
\end{tabular}
\end{center}

In this election, $a$ will be the winner whenever $l \leq k$.
Otherwise the winner will be $b$.

\vspace{1em}

\noindent
$(\{b,w\},V')$

\begin{center}
\begin{tabular}{c|c}
  Candidate & Score \\
  \hline
  $b$         & $6nr + 6nmr + 2(m-k-1)nr + 4r$ \\
  $w$         & $4nr + 6nmr + 2(m-k-1)nr + 2 s_0(w)$\\
\end{tabular}
\end{center}

In this case, $b$ is the clear winner.

\vspace{1em}

\noindent
$(\{a,w\} \cup D, V')$ where $D \subseteq C^*, |D| = l$

\begin{center}
\begin{tabular}{c|c}
  Candidate & Score \\
  \hline
  $c \in D$ & $4nr + 6nmr + 2(m-k-1)nr + 2 s_0(c)$ \\
  $a$ & $4nr + 4nmr + 2(m-l)nr + 2$ \\
  $w$ & $4nr + 6nmr + 2(m-k-1)nr + 2 s_0(w)$\\
\end{tabular}
\end{center}

The winner will again be whatever candidate is the winner over the
original voter set $V$.

\vspace{1em}


We can show that if $w$ can be made the winner of $(C,V)$ by
deleting $\leq k$ candidates, $w$ can be made the winner of $(C', V')$
through control by runoff partition of candidates.
Suppose $w$ can be made the winner of $(C,V)$ through deleting $\leq
k$ candidates.  Let $D$ be the set of candidates which were deleted in
the deletion problem.  Partition the candidates into the subelections
$(D \cup \{a,b\}, V')$ and $(C-D, V')$.  $a$ will win the first
subelection as shown above.  $w$ will win the second subelection, as
it must if it is capable of winning with the candidates in $D$
deleted.  The final election will then come down to $w$ and $a$, and
as we see above, $w$ will come out the victor.  Alternately, in the
non-runoff partition case, let the initial subelection be $(D \cup
\{a,b\}, V')$, which $a$ will win.  The final election will come down
to $(\{a,w\} \cup C-D, V')$, which $w$ will win.

We can show that if $w$ can be made the winner of $(C', V')$ through
control by runoff partition of candidates, $w$ can be made the winner
of $(C,V)$ by deleting $\leq k$ candidates.  Suppose $w$ can be made
the winner of the election $(C', V')$ through control by runoff
partition of candidates.  It must be that this occurs through a
partition of the form $(\{a.b\} \cup D, \{w\} \cup (C^* - D))$ with $D
\subseteq C^*, |D| \leq k$.  $b$ will always beat $w$, so they cannot
face each other in either the initial or final elections.  The only
candidate capable of beating $b$ is $a$ when not in an election with
$w$ and when accompanied by no more than $k$ other candidates from
$C$. $w$ must also be able to defeat the remaining $m-k$ candidates
from the initial election. Consequently $w$ can also be made the
winner of $(C,V)$ by deleting $k$ candidates. $\Box$

\vspace{2em}

The preceding construction will shows that NRV is resistant to
constructive cases of partition of candidates.  However it is not
sufficient for the destructive cases, as a winning candidate in the
original election $(C,V)$ will not actually win in $(C', V')$.
Thus we will present a new construction to handle the
destructive cases.

\begin{theorem}
2-NRV is resistant to destructive control by partition and runoff
partition of candidates.
\end{theorem}

\paragraph{Proof}
We can reduce the hitting set problem to the problem of destructive
control by partition of candidates.  Let $(B, {\cal S}, k)$ be an
instance of hitting set where $B = \{b_1, b_2, \ldots, b_n\}$, ${\cal
  S} = \{S_1, S_2, \ldots, S_m\}, S_i \subseteq B$, and $k \in
\mathbb{N}^+, k \leq n$.

We will construct a $2$-range election based on this instance.
The candidate set $C$ will consist of $B \cup \{w\}$.  The voter set $V$ will
be as follows.

\begin{itemize}
\item For each $S \in {\cal S}$, $4(k+1)$ voters have a score of $2$ for each $b
  \in S$ and a score $1$ for $w$
\item For each $S \in {\cal S}$, $4(k+1)$ voters have a score of $2$ for each $b
  \in B, b \notin S$, and a score of 0 for every other candidate.
\item For each $b \in B$, $4$ voters have a score of $2$ for $b$, a
  score of 1 for each $b' \in B, b' \neq b$, and a score of 0 for $w$
\item $2(k+1)m + 4n - 2k + 1$ voters have a score $2$ for $w$ and a
  score of 0 for every other candidate.
\end{itemize}

Again, we will first consider the outcome of several forms of subelections of
this election.

\vspace{1em}

\noindent
$(C,V)$

\begin{center}
\begin{tabular}{c|c}
  Candidate & Score \\
  \hline
  $w$ & $8(k+1)m + 8n - 4k + 2$  \\
  $b \in B$ & $8(k+1)m + 4n + 4$ \\
\end{tabular}
\end{center}

$w$ will win this election for any $k \leq n$.  

\vspace{1em}

\noindent
$(\{w\} \cup D,V)$, $D$ is a hitting set, $|D| = l$

\begin{center}
\begin{tabular}{c|c}
  Candidate & Score \\
  \hline
  $w$ & $8(k+1)m + 8n - 4k + 2$  \\
  $b \in B$ & $8(k+1)m + 8n-4l+4$ \\
\end{tabular}
\end{center}

If $l \leq k$, every $b \in B$ will tie for first with $w$ as the
clear loser.  Otherwise $w$ will be the winner.  

\vspace{1em}

\noindent
$(\{w\} \cup D,V)$, $D$ is not a hitting set, $|D| = l$

\begin{center}
\begin{tabular}{c|c}
  Candidate & Score \\
  \hline
  $w$ & $8(k+1)m + 8n - 4k + 2 + 4(k+1)$\\
  $b \in B$ & $8(k+1)m + 8n - 4l + 4$ \\
\end{tabular}
\end{center}

$w$ will win this election.  

\vspace{1em}

There is a hitting set $B' \subset B$ where $|B'| \leq k$ if and
only if $w$ can be made to lose the election through partition or
run-off partition of candidates.

If there is a hitting set $B' \subset B$ of size $\leq k$, $w$
can be made to lose the election through control by partition or
runoff partition of candidates with the partitions $\{w\} \cup B',
B-B'$. $w$ will lose the initial subelection $(\{w\} \cup B')$ as shown
above, and will thus lose the entire election.

If there is no such hitting set, $w$ cannot be made to lose the
election through partition or runoff partition of candidates.  As
shown above, $w$ will win any election $(\{w\} \cup B', V)$ where
$|B'| > k$, or where his or her opponents do not comprise a hitting
set.  The one special case is when $k=n$, where $w$ will lose the
original election but in this case there is a trivially always a
hitting set, so this problem should also always accept.  In any other
case $w$ will win against any subset of $B$, so $w$ will win both any
initial subelection and the final election, and so there is no
partition to make them lose. $\Box$

\section{Conclusions}


This work leaves open a number of questions.  NRV still falls
short of resistance to all cases of control, so some other natural
system could still best it.  Just as useful would be results about the
conditions that are required for a voting system to have various
resistances.  It may still be that natural systems are incapable of
having every control resistance simultaneously.  Any useful results
here would first require a formalization of what exactly a natural
voting system is.  Most desirable would be a reasonable set of
conditions that could be shown to be incompatible with holding all
resistances simultaneously, \`{a} la Arrow's Theorem.

Other useful work would be to analyze methods for sidestepping the
worst-case difficulty of the control problems here.  One example is the
use of approximation algorithms as studied by Brelsford et
al.~\cite{bre-fal-hem-sch-sch:c:approximating-elections}, or analysis
of the problems with a restricted preference-ensemble
model~\cite{fal-hem-hem-rot:j:single-peaked-preferences}.  It is
important to note that the worst-case analysis performed here does not
provide a guarantee that the problems will be hard on average.  Still,
this work provides a good first step toward understanding the behavior
of these systems with respect to control.

\section*{Acknowledgments} 
For helpful comments and suggestions, I am grateful to Edith
Hemaspaandra, who advised my M.S\@. thesis in which an earlier version
of part of this work appeared, Lane Hemaspaandra, Preetjot Singh, and
Andrew Lin.

\bibliography{grycurtis}
\bibliographystyle{alpha}

\end{document}